# Saturation of Spontaneous Polarization Charge in Pyroelectric Crystals of $LiNbO_3$, $LiTaO_3$ and $CsNO_3$ at Low Temperature Above 4.2 K

James D. Brownridge* and Sol Raboy
Department of Physics, Applied Physics and Astronomy
Binghamton University, P.O. Box 6000 Binghamton, New York 13902-6000

## Abstract

Experimental observations of the change in the polarization charge of pyroelectric crystals of $LiNbO_3$, $LiTaO_3$ and $CsNO_3$, as the temperature of the crystal is changed from about 300K to a lower limit of 4.2K, are described. It was found that the rate of change of the polarization charge slowed down considerably and, perhaps, reached a zero rate of change. The onset of this "saturation" of the polarization charge occurred at low temperatures, above the temperature of liquid helium, which were different for each of the three types of crystal.

## INTRODUCTION

In a paper by R. S. Weiss and T. K. Gaylord[1] the increase in the magnitude of the polarization charge as the temperature of the pyroelectric crystal of $LiNbO_3$ decreased was associated with the increase of the average displacement of the atoms of Lithium and Niobium out of the plane of the atoms of Oxygen. It was reasonable to expect that the average displacement of the atoms of Li and Nb would ultimately reach a maximum limiting value. The authors pointed out that their reasoning also applied to $LiTaO_3$, which has a crystal structure similar to that of $LiNbO_3$. Hence, a series of experiments were undertaken to explore the possibility that at some low temperature the polarization charge would stop increasing in magnitude as the temperature of the crystals decreased. The investigations were extended to include crystals of $CsNO_3$, which were grown in this laboratory. In a testimonial paper[2] "On the Quantum Theory of Pyroelectricity" Max Born summarized his investigations of the low temperature behavior of pyroelectric crystals. With certain premises and approximations this pioneer work came up with a $T^2$ dependence of the electric moment at low temperatures. This result was in agreement with the then existing experimental evidence down to 23K. Since Born's paper there have been publications[3-12] reporting experiments of different crystals down to temperatures of liquid helium and below. None of these papers were concerned with the attainment of "saturation" of the polarization charge.

In the experiments described below, the changes in the polarization charge, as the temperature of the crystals was lowered to 4.2K, were measured. The experiments consisted of measurements of the polarization charge with sensitivities of hundredths of nanocoulombs and measurements of the polarization currents with sensitivities of hundredths of picoamperes. Both sets of measurements were made in the presence of background effects, the bulk of which mimicked the motion of the crystal assembly. The final results, as presented in Table I, were calculated from data of the measurements of the polarization currents. The term "polarization charge", that is used in the descriptions that follow, will refer to the charge appearing at the surface of the crystal and is fixed in the crystal, which changes in magnitude as the temperature changes. The term "polarization current" refers to the current that ensues as the polarization charge changes in magnitude. The term “onset” or “onset temperature” will refer to the temperature, during the cooling process, at which the rate of change of the polarization charge changes drastically to a smaller value and perhaps to zero. The term “exit” or “exit temperature” will refer to the temperature, during the warming process, at which the rate of change of the polarization charge will suddenly increase dramatically from the slowly varying change or possibly the constant value, i.e., “saturation” as the temperature of the crystal is increased.

## EXPERIMENTAL PROGRAM

The early investigations of the change in the polarization charge of crystals of $LiNbO_3$, $LiTaO_3$ and $CsNO_3$, as the temperature of the crystals was varied from about 300K to 4.2K and back to about 300K, consisted of a series of experiments, in which the change in the polarization charge relative to the polarization charge at about 300K was measured. The experimental arrangement, for the early experiments and subsequent experiments, is shown in the schematic diagram, Fig. 1. The crystal assembly was contained in a hollow tube, which was driven by a motorized pulley. The temperature of the crystal assembly was changed by moving the assembly from immersion in the liquid helium, 4.2K, to positions above the surface of the liquid, i.e., to higher temperatures. By selecting the positions of two micro switches, equipped with magnetic sensors, desired temperature intervals within the limits could be selected. The micro switches, when actuated by the magnetic sensors stopped the electric current to the motor, i.e., halted the motion of the crystal assembly, and then started the current to the driving motor in the reversed direction, i.e., moved the crystal assembly in the opposite direction. Thus, with the micro switches in selected positions the crystal assembly moved in cyclical manner through established temperature limits.

The crystal under study is represented by block A in Fig. 2, the crystal assembly. There were two leads attached to the crystal, one to the +z base and one to the –z base. With two electrometers it was possible to study the change in polarization charge with temperature change at the two bases simultaneously. During a large part of the time of the investigation, one base was grounded and one base was connected to an electrometer, Keithly Model 610C. The temperature of the crystal under study was measured by a temperature sensor, CY7 series, from Omega Engineering. The temperature sensor is represented by block B in Fig. 2. The manufacturers quote an accuracy of 0.25K in the range 2K to 100K and an accuracy of 0.50K in the range 100K to 400K.The two conducting leads attached to the crystal and the four leads from the temperature measuring device were brought out through a hollow tube of stainless steel to the appropriate recording device. The two leads to the crystal were attached to the crystal with a very thin film of silver conducting epoxy. The temperature sensor was fastened to the crystal with pliable thermally conducting paste, Weldwood “Contact Cement”. Thus, it was assumed that the crystal was essentially free during the acquisition of data. During the experimental process the crystal was completely immersed either in liquid helium or in the gaseous helium above the liquid. The entire surface of the crystal, therefore, was at a uniform temperature throughout the series of experiments, thereby avoiding the problem of temperature gradients between the +z base and the -z base of the crystal.[13]

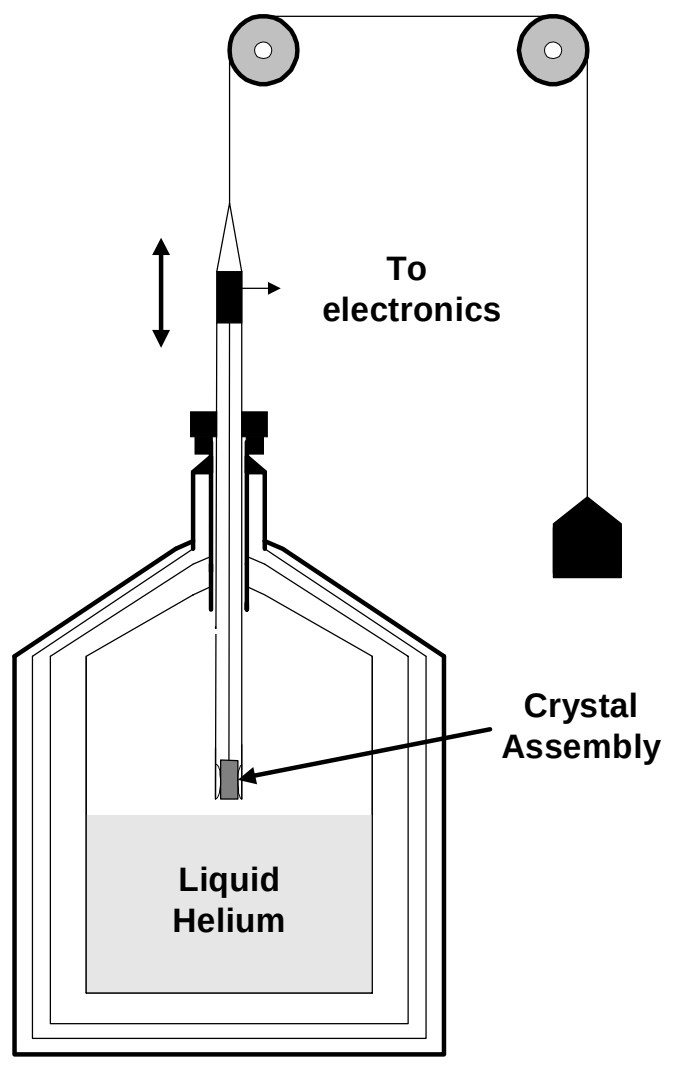

Figure 1: Schematic diagram of the apparatus used to vary the temperature, for the study of the changes in the polarization charge, of the crystals under study. The crystal assembly was driven by a motorized pulley, so that the crystal could be moved into and out of the liquid helium, 4.2K, to higher positions in the Dewar, up to about 300K. By setting the positions of two micro switches, with attached magnetic sensors, along a vertical axis external to the Dewar, any desired interval of temperature between 4.2K and 300K was attainable. The micro switches were used to reverse the direction of the motion of the crystal assembly and thus a series of thermal cycles for a prearranged interval of temperature could be applied to the crystal.

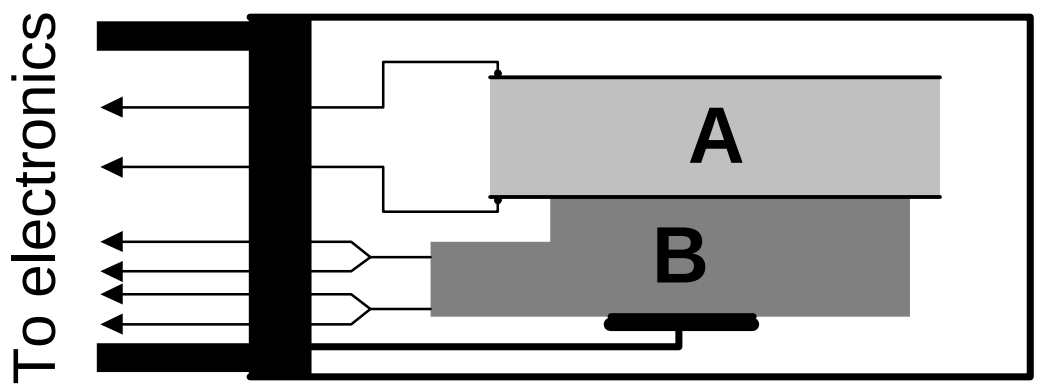

Figure 2: Schematic diagram of the crystal assembly. Block A represents the crystal under study with the two leads from the electrometer, one lead to the +z base and one lead to the –z base. Block B represents the temperature sensor, CY 7 from the OMEGA Company. The crystal and temperature sensor were sealed in a vacuum tight enclosure.

Two electrometers, in the charge mode, were used to measure the change in the polarization charge with respect to the polarization charge at about 300K at the +z base and the-z base of the crystal under study simultaneously from about 300K to 4.2K and back to about 300K. Data for crystals of $LiNbO_3$, $LiTaO_3$ and $CsNO_3$ were obtained. The crystals of $LiNbO_3$, $LiTaO_3$ were obtained from Crystal Technology, Inc., Palo Alto, CA. At the start of the experiments, the probe circuits of the electrometers were grounded at the crystal temperature of about 300K, i.e., the electrometers were zeroed.[14] Although the electrometers read zero, there were induced charges in the probe circuits equal and opposite in sign to the polarization charge on the surfaces of the crystal at that temperature[14]. The preceding statement presumes that the temperature is below the Curie temperature of the crystal. The grounds were broken and data was accumulated at intervals of one second as the temperature of the crystals was lowered to 4.2K by lowering the crystal assembly slowly through the gaseous space to immersion in the liquid helium. The crystal assembly was then raised slowly out of the liquid helium and then through the gaseous space until a temperature of about 300K was reached.

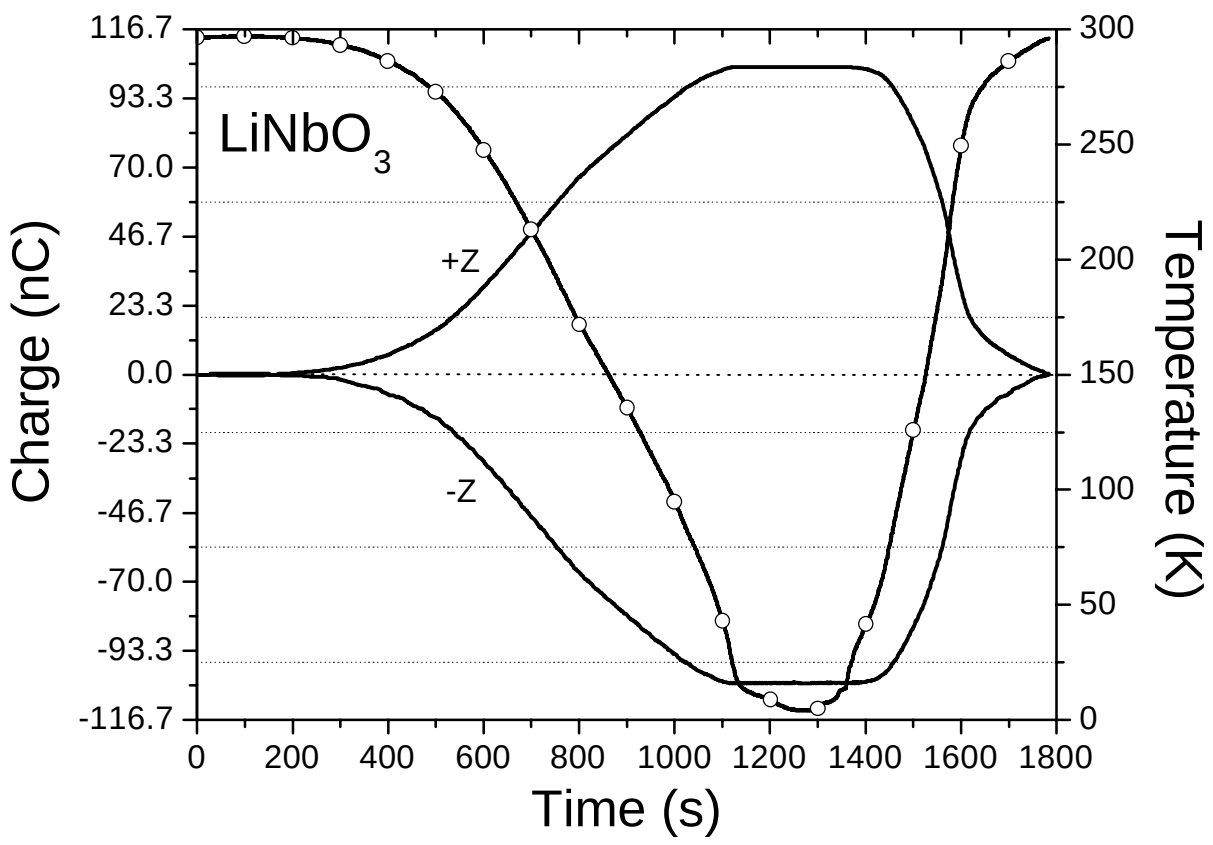

Figure 3: Temperature of the crystal and the change in polarization charge from the polarization charge at 300K of a crystal of $LiNbO_3$ as functions of time. The curve with the open circles is a plot of the temperature as a function of time. The solid unmarked curves represent the change in the polarization charge relative to that at 300K at the +z base and the-z base of the crystal. The data for the three curves were recorded simultaneously

The results of an experiment on a crystal of $LiNbO_3$, base area of (10.19±0.04mm$^2$) by a thickness of (1.001±0.001mm, are presented in Fig. 3. Similar curves were obtained for crystals of $LiTaO_3$ and $CsNO_3$. The charge readings, at the temperatures indicated in Fig. 3, are actually the change in the polarization charge from the polarization charge at a crystal temperature of about 300K. The representation in Fig. 3 shows that, as the temperature was lowered, the polarization charge at the bases of the crystals increased numerically, with positive sign on the +z base and negative sign on the –z base. The representation shows that at a very low temperature the rate of change of the polarization charge with change in temperature slows down considerably and perhaps levels off. In all three crystal types the onset of this very slow rate of change, very definitely, appears at a temperature above 4.2K, that of liquid helium. As the temperature was raised slowly from 4.2K the amount of polarization charge appeared to be constant until about 36K. Then the amount of polarization charge started to decrease. The rate of decrease increased as the temperature increased until at about 300K the initial starting conditions were achieved. The rate of change with temperature of the polarization charge below about 36K relative to the charge at 300K was very small compared to the actual value of the relative charge, about 115nC. The electrometer sensitivity used to acquire the data for Fig. 3, therefore, was not suitable for the precise determination of the temperature at which the relative polarization charge became “constant”. Further investigation was, therefore, indicated.

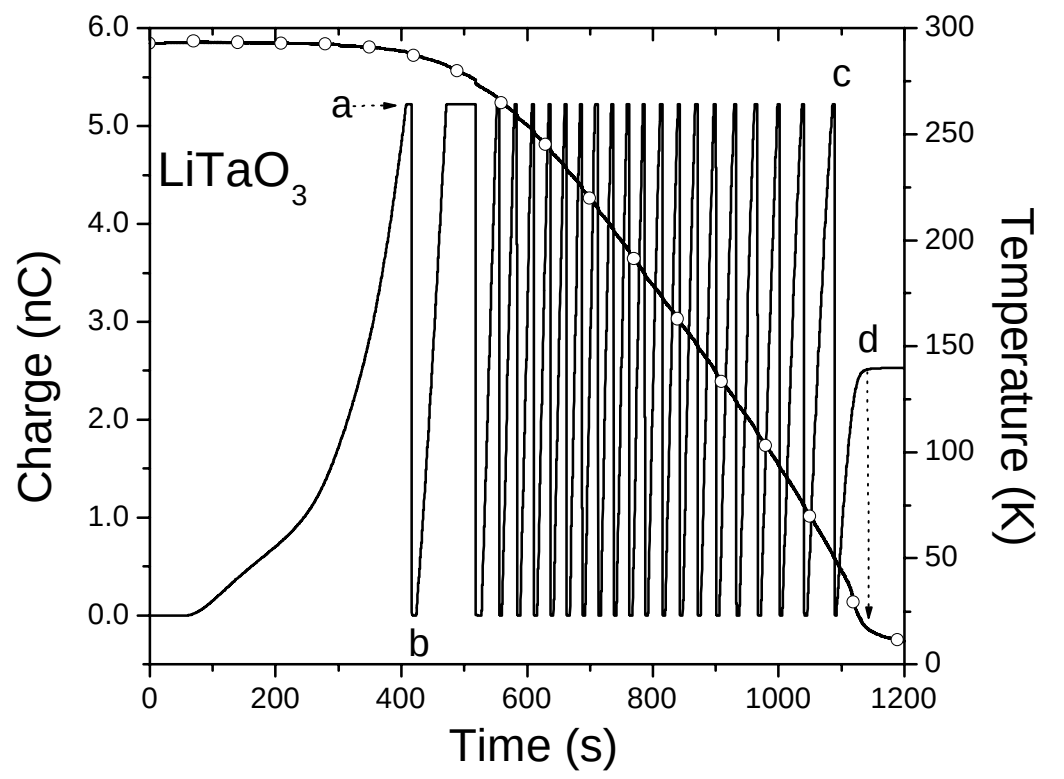

Figure 4A: Temperature of the crystal and the polarization charge, at high charge sensitivity, relative to the polarization charge at selected temperatures of a crystal of $LiTaO_3$, as functions of time. The curve with the open circles represents the temperature of the crystal as a function of time. The solid unmarked curve represents the relative polarization charge at the +z base of the crystal measured with high charge sensitivity. Because of the high charge sensitivity of the electrometer used for the accumulation of this data, the electrometer-computer system pegs, i.e., goes off scale at a reading of about 5nanocoul. At the start of the accumulation of data at 300K the input circuit of the electrometer was grounded, i.e., the electrometer was zeroed. Then between 275K and 300K the charge measuring system pegged, shown at point (a). The input circuit was grounded at this point; the zero reading of the electrometer is indicated at point (b). The subsequent readings of the electrometer, until the electrometer pegs again, are now the change in the polarization charge of the crystal from that at the temperature at which the crystal was grounded. This process was repeated each time the electrometer system went off scale There were twenty more pegs of the electrometer system, exclusive of the one at point (a), each followed by the grounding of the electrometer input circuit and then a continuation of the accumulation of data. After the peg at point (c) and the grounding of the electrometer that followed The rate of change of the relative charge slowed down after the temperature corresponding to point (d) was reached and passed.

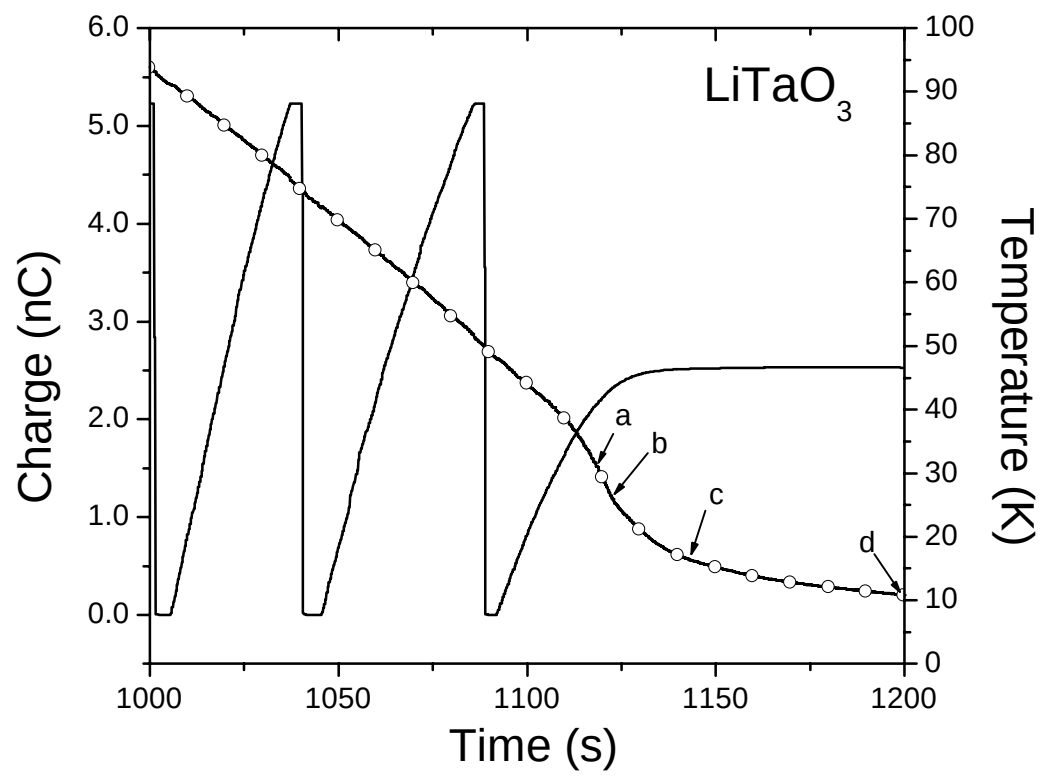

Figure 4B: Expansion of the plots of $LiTaO_3$ around the region of point (c) to include the region beyond point (d) of Fig. 4A, i.e., the interval of time from 1,000s to 1,200s. From point (a) to point (b) the temperature changed from 31.52K to 26.27K and the relative polarization charge changed from 2.12nC to 2.45nC. From point (c) to point (d) the temperature changed from 16.70K to 11.02K and the relative polarization charge changed from 2.52nC to 2.53nC

With a more sensitive electrometer scale, the experiments were repeated. Initially the probe circuit of the electrometer was grounded at a crystal temperature of about 300K. As the temperature of the crystal was lowered through a small range from about 300K the electrometer went off scale, because of the much greater sensitivity of the measuring apparatus. The electrometer probe circuit was then grounded, i.e., zeroed as the temperature continued to decrease. As the temperature of the crystal continued to decrease, the electrometer readings indicated an increase in the amount of polarization charge relative to the polarization charge at the temperature of the previous grounding[14]. As the experiment continued, the temperature continued to decrease, the amount of relative polarization charge increased until the electrometer went off scale again. The grounding procedure was repeated as the temperature continued to decrease until the electrometer pegged again. The grounding procedure was applied again. This technique was repeated until the charge reading started to change very slowly and appeared to level off at a constant value. This occurred at a crystal temperature below about 15K for each of the crystals examined in this experiment. The results of the experiment at this high sensitivity on the +z base of a crystal of $LiTaO_3$, with a base ($10.54\pm0.07mm^2$) and a thickness (1.00±0.00mm,) are presented in Fig. 4A. At the start of the accumulation of data, the probe circuit of the electrometer was grounded at about 300K. After the ground was broken the temperature was lowered by moving the crystal assembly slowly down through the gas of the Dewar to immersion in the liquid helium. The electrometer reading of the relative polarization charge and the reading of the temperature sensor were recorded every second in the computer. Point (a) in Fig. 4A indicates the first time the electrometer went off scale and point (b) indicates the subsequent grounding of the electrometer so that the electrometer reads zero. The presentation in Fig. 4A shows the subsequent times when the electrometer pegged followed by zeroing and then continuation. Point (c) is the last time the electrometer pegged followed by the last time the electrometer was zeroed. After a temperature corresponding to point (d), the charge reading appeared to change very slowly with temperature change, and perhaps level off to a constant value. The horizontal scale, 1200sec, used in Fig. 4A does not allow for a precise determination of the temperature at which the polarization charge of the crystal becomes constant. An expanded scale, therefore, of the last 200 seconds of Fig. 4A, presented as Fig. 4B was used to identify the region of the data where constancy of the polarization charge might be seen. The actual data, i.e., the relative polarization charge and the temperature, corresponding to points (a), (b), (c) and (d) on Fig. 4B were examined to explore the onset of constancy of the polarization charge. From point (a) to point (b) The temperature changed from 31.52K to 26.27K while the relative polarization charge changed from 2.12nC to 2.45nC. On the other hand from point (c) to point (d) the temperature changed from 16.7 K to 11.02K while the polarization charge changed from 2.52nC to 2.53nC. The indication is that somewhere below about 16K there appeared to be a decrease, perhaps to zero, in the rate of change of the polarization charge of the crystal of $LiTaO_3$ as the temperature was lowered.

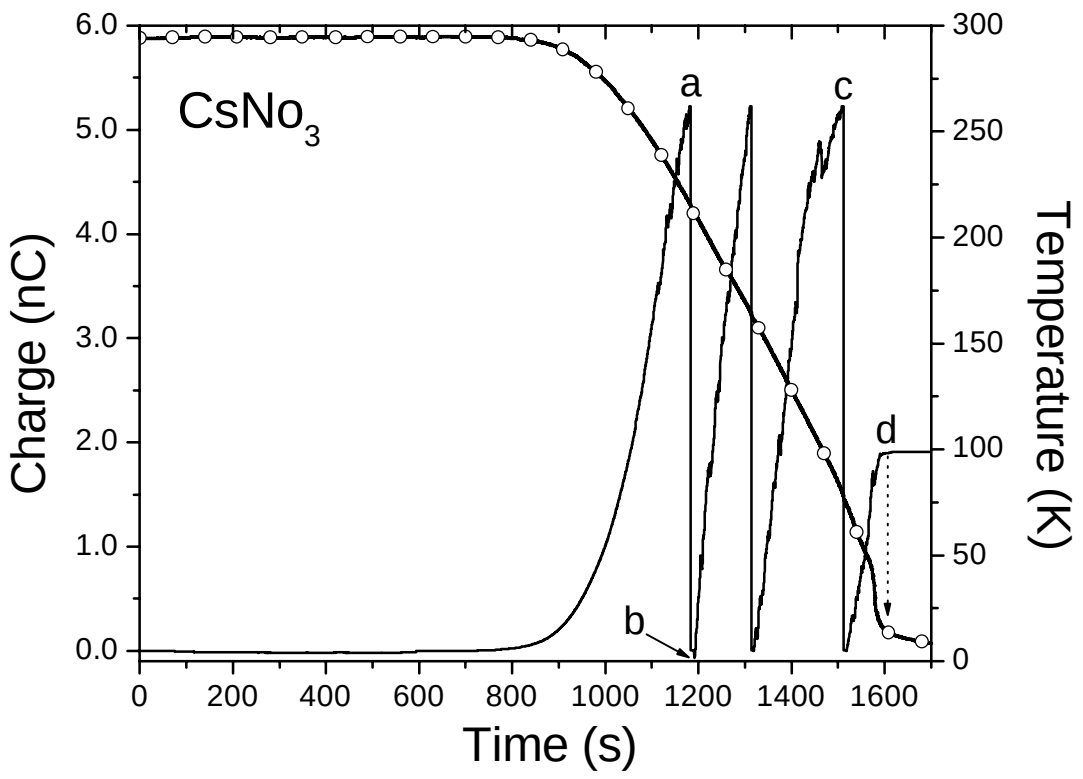

Figure 5A: Temperature and relative polarization charge at the +z base of a crystal of $CsNO_3$ as functions of time. The curve with the open circles is the plot of the temperature of the crystal as a function of time. The solid unmarked curve is the plot of the change in the polarization charge from that of the polarization charge at the temperature of the just prior grounding of the input circuit of the electrometer. This data was taken at high charge sensitivity. Initially the probe circuit at the input to the electrometer was grounded at a crystal temperature of about 300K. The temperature corresponding to point (a) is the temperature for which the electrometer-computer system went off scale for the first time. The probe circuit was grounded at this temperature as is indicated at point (b). During the subsequent lowering of the temperature the amount of charge increased sufficiently to send the recording system off scale two more times and therefore each peg was followed by a grounding of the probe circuit. The last peg, before a temperature of 4.2K was reached, is indicated by point (c). Before the temperature of 4.2K was reached the rate of change of the relative polarization charge became very small almost to zero from point (d) on, suggesting further study.

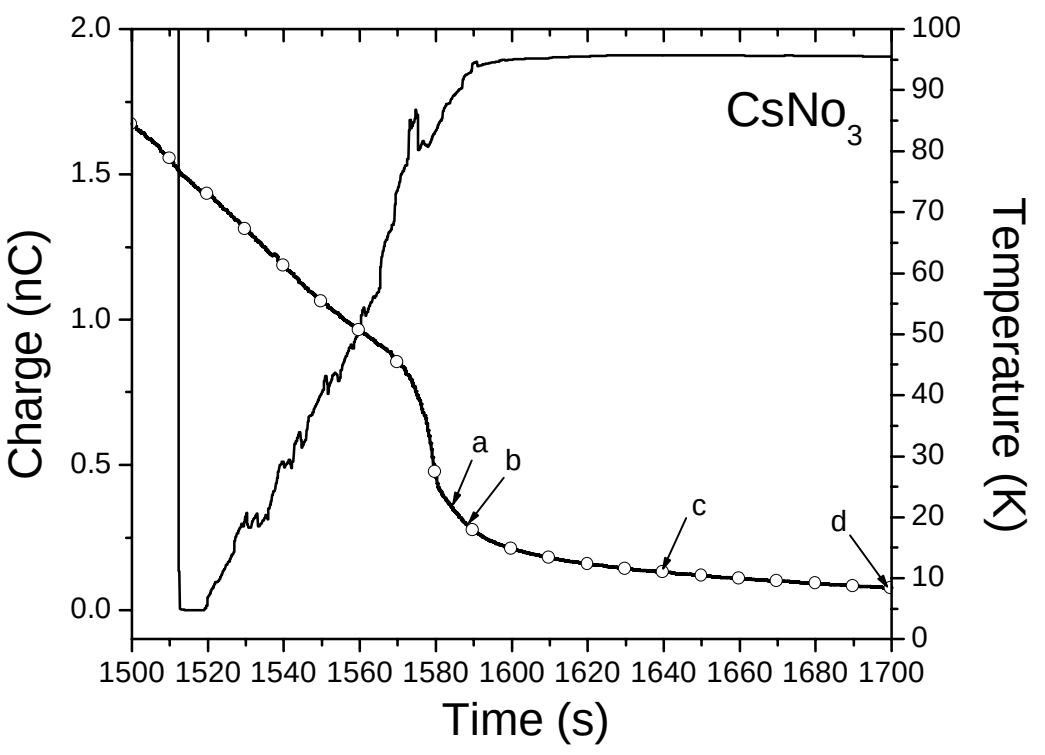

Figure 5B: Expansion of part of Fig. 5A corresponding to the interval of time of 1500s to 1700s. From point (a) to point (b) the temperature changed from 22.04K to 19.32K. The relative polarization charge changed from 1.74nC to 1.83nC. From point (c) to point (d) the temperature changed from 11.54K to 8.76K and the relative polarization charge, to three significant places, remained constant at 1.91nC.

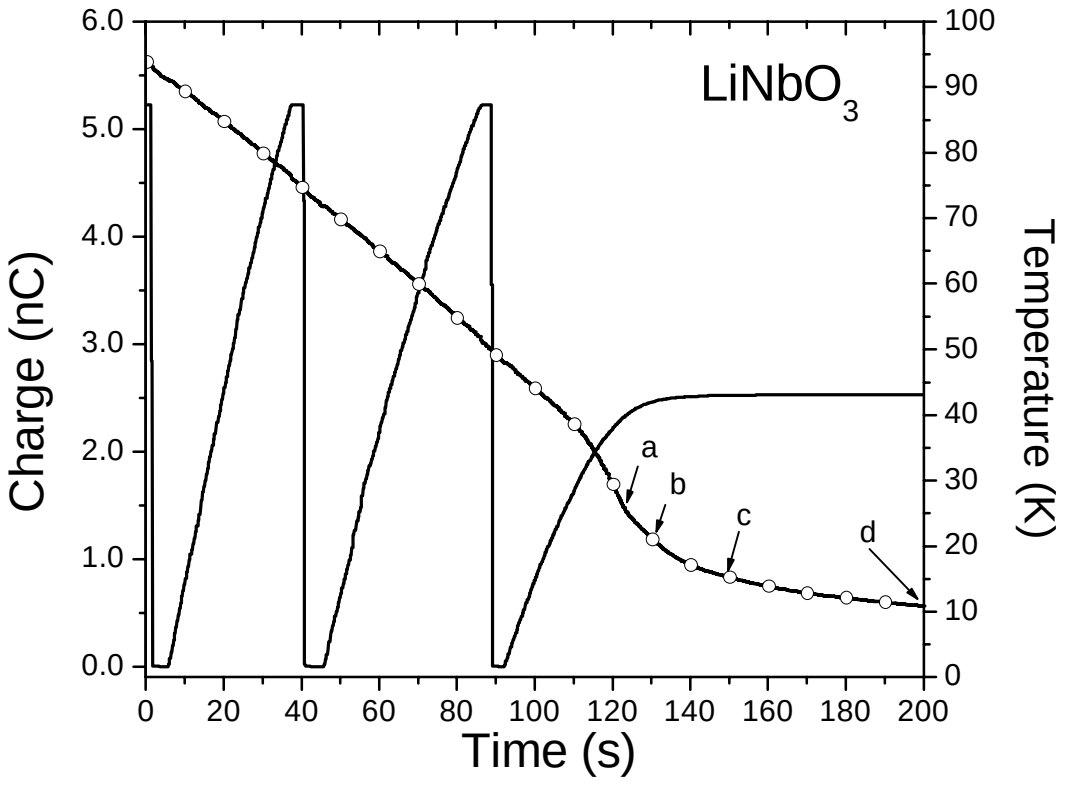

Figure 6: Expanded part, about 100K to about 16K, of temperature and relative polarization charge at high sensitivity of a crystal of $LiNbO_3$ as functions of time. The curve with the open circles represents the temperature as a function of time. The solid unmarked curve represents the relative polarization charge as a function of time, i.e., relative to the polarization charge at the temperature of the preceding grounding of the probe circuit of the electrometer. From point (a) to point (b) the temperature changed from 26.1K to 21.2k while the polarization charge changed from 2.310nC to 2.453nC. From point (c) to point (d) the temperature changed from 15.8K to 11.0K while the relative polarization charge, to four significant places, appeared steady at 2.517nC.

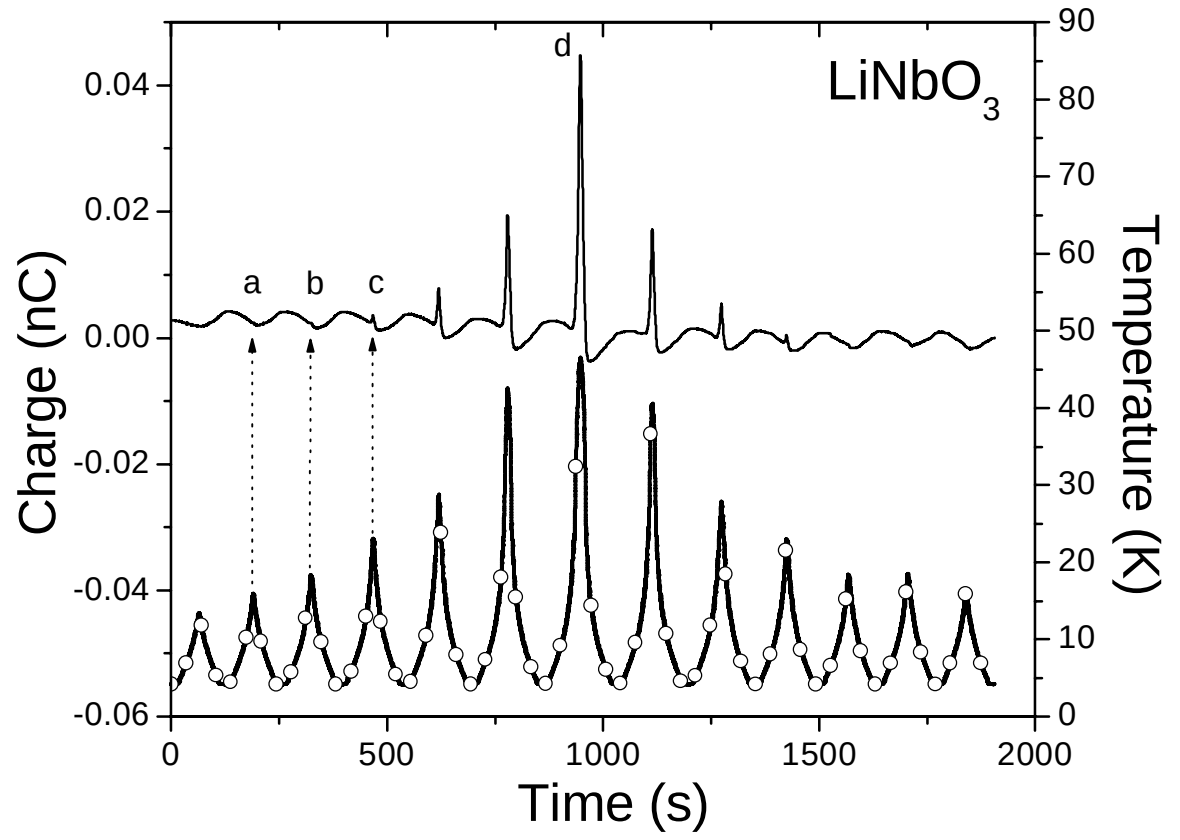

Figure 7A: Temperature and the change in the polarization charge from the polarization charge at 4.2K at the –z base of a crystal of $LiNbO_3$ as functions of time. The curve with the open circles represents the temperature as a function of time obtained simultaneously with the unmarked solid curve, which represents the difference in the polarization charge from that at 4.2K as a function of time. The measurements of the polarization charge were made with high sensitivity. The spikes appearing on the wavelike variations of the charge measurements are attributed to the change in the polarization charge and the wavelike part of the charge measurements are attributed to "pick up" resulting from the motion of the crystal assembly. For the temperature swing associated with point (a) the charge response is smooth. The effect of a slightly larger temperature swing are seen in the charge response at point (b) as a break in the decreasing side of the wave. As seen in point (c) a small peak appears in the wave for a larger swing in the temperature. As the maximum temperature of each swing is increased, the spike on the decreasing side of the wave increases. The spike at point (d) is associated with the largest temperature swing of the figure.

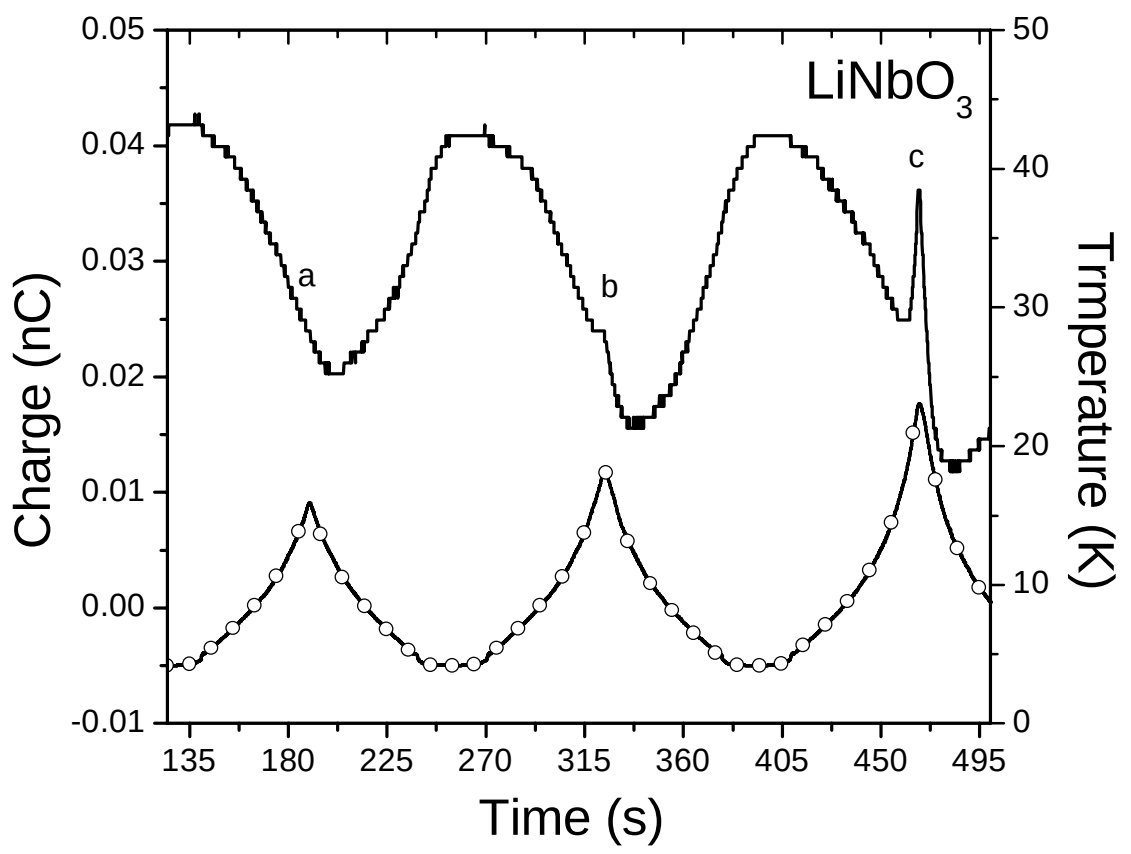

Figure 7B: Expansion of the region of the plots of Fig. 7A for the time interval of about 135s to about 495s. It is seen that for the temperature swing corresponding to point (a) there is no break in the decreasing side of the "pick up" cycle. But with a slight increase in the temperature swing a break appears in the decreasing side of the "pick up" cycle at point (b). With an additional increase in the temperature swing the break in the "pick up" cycle becomes a spike. The spikes in the subsequent cycles establish the signature of the exit of the crystal from "saturation" as the temperature increased.

Similar data, at the high sensitivity of the electrometer in the charge mode, was obtained for a crystal of $CsNO_3$, with a base area ($9.319\pm0.021mm^2$) and a thickness ($3.012\pm0.001mm$). The crystals of $CsNO_3$ were produced in this laboratory and had pronounced mosaic structures. The results of this experiment are exhibited in Fig. 5A. The experiment started off with grounding, i.e., zeroing the electrometer at about 300K. At point (a) the electrometer went off scale and point (b) shows the result of the grounding operation. Point (c) indicates the last time the electrometer pegged which is followed by the final grounding of the probe circuit. Point (d) corresponds to the temperature at which the change in the charge reading with temperature change became appreciably smaller. The plot shows some jagged peaks, which are attributed to electrical discharges because of the lack of perfection of the crystals of $CsNO_3$ as the amount of polarization charge increases numerically with decreasing temperature. The plot in Fig. 5A of the polarization charge and the temperature of the crystal as functions of time was presented for 1200 seconds. It is not clear from this plot that there is any leveling off of the relative polarization charge as the temperature of the crystal fell to low values. An expansion of the last 200 seconds of the plot, Fig. 5B, however, suggests a region of the data to be examined. For example, from point (a) to point (b) the temperature of the crystal changed from 22.04K to 19.32K while the polarization charge changed from 1.74nC to 1.83nC. On the other hand from point (c) to point (d) the temperature changed from 11.54K to 8.76K while the relative polarization charge, for three significant places, remained constant at 1.91nC. These observations lead to an inference for crystals of $CsNO_3$, that the rate of change of polarization charge with temperature becomes very small or may approach zero at a low temperature.

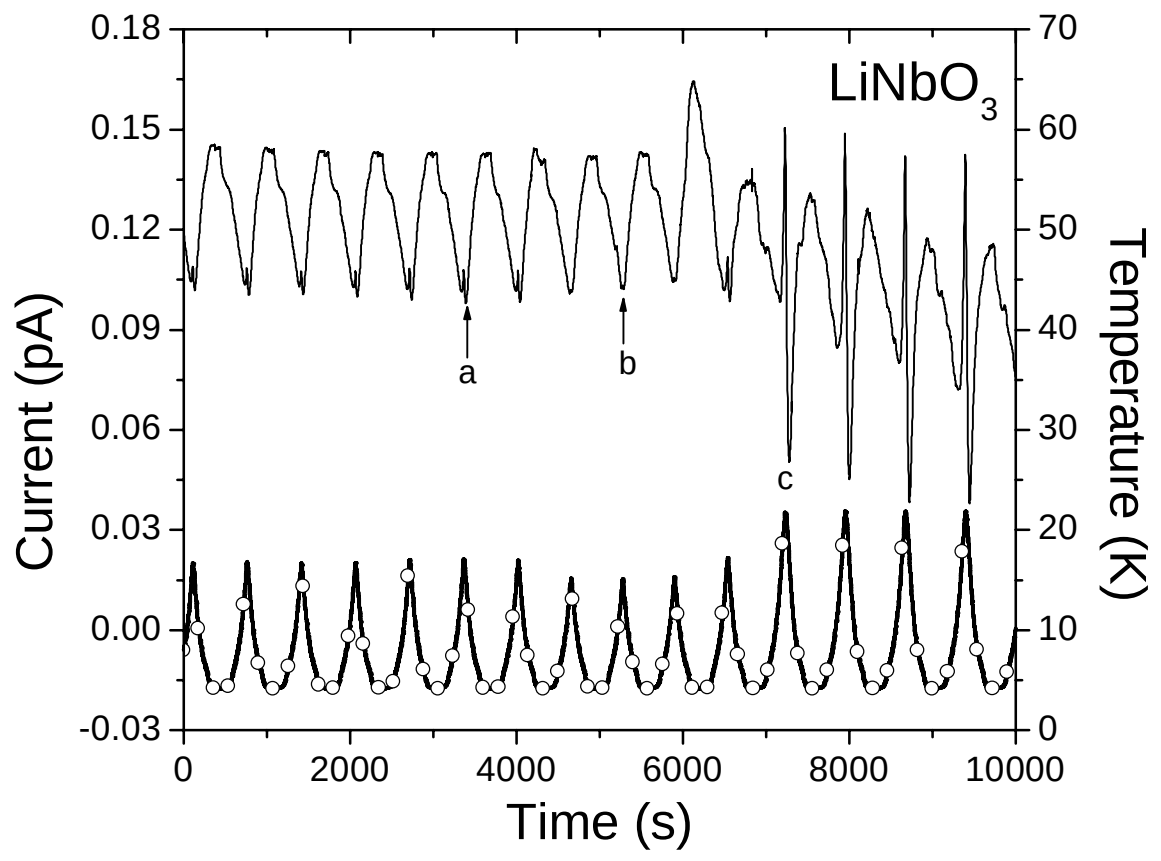

Figure 8A: Temperature and polarization current at the –z base of a crystal of $LiNbO_3$ as functions of time. The data for the temperature and the polarization current were obtained simultaneously. The curve with the open circles is the temperature plotted as a function of time. The unmarked solid curve represents the polarization current plotted as a function of time. Again there appears a cyclical pattern in the current structure, which follows the motion of the crystal assembly, which will be designated as the "pick up" current. A small spike, as at point (a), is observed. For a slightly smaller temperature swing the spike is imperceptible, as at point (b). The break develops into a large peak, as seen at point (c), for the large temperature swing.

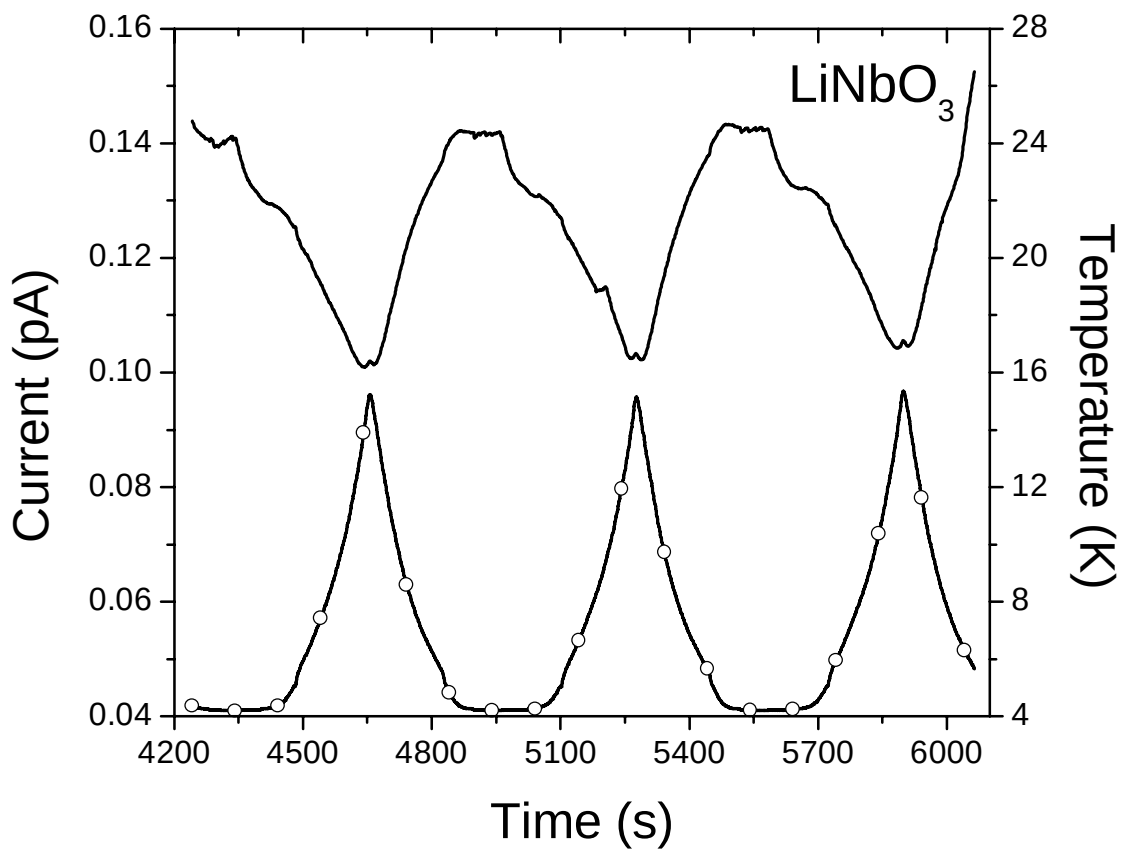

Figure 8B: Expansion of the response of $LiNbO_3$ in the time interval of about 4,200s to about 6,000s, which shows the polarization current centered at point (b) of Fig. 8A, corresponding to the three smallest temperature cycles. At this large magnification, the breaks in the current are clearly visible.

Crystals of $LiNbO_3$ were studied in similar fashion i.e., data for the temperature and the polarization charge relative to that at about 300K as functions of time were accumulated for a range of about 300K to 4.2K. The plots of the data show points, equivalent to the point (d) of Figs 4A and 5A, where the rate of change of the polarization charge appear to slow down to almost zero. A typical plot of the final part of the temperature range, from about 100K to about 16K is presented in Fig. 6. From point (a) to point (b) the temperature changed from 26.1K to 21.2K while the relative polarization charge changed from 2.31nC to 2.45nC. Whereas, from point (c) to point (d), where the temperature changed from 15.8K to11.0K, the relative polarization charge remained steady to four significant places at 2.517nC.

In order to ascertain, more precisely, the temperature of the crystal at which "saturation" starts, measurements of the change in the polarization charge relative to the polarization charge at the crystal temperature of liquid helium were undertaken. The probe circuit of the electrometer was grounded while the crystal assembly was immersed in the liquid helium. The assembly was raised slowly to slightly higher temperatures up to a pre-selected maximum, after the ground was broken, and then lowered to immersion in the liquid helium. The results of the experiment at the -z base of a crystal of $LiNbO_3$ are shown in Figs. 7A and 7B.. The maximum of the each temperature swing was successively increased from about 13K to about 45K and then decreased successively to 17K. As the temperature maximum of each cycle was increased, the charge appeared to have a sudden peak on the decreasing side of a cyclical part of the response, the larger the maximum of the temperature cycle the larger the sharp peak. Both of these contributions to the charge reading appeared to be riding on a slowly changing component of the charge. The identification of the contribution of the polarization charge to the charge reading was made by the accumulation of results for temperature swings below about 13K followed by swings to higher maxima. A typical response to the small temperature swings, from 4.2K to about 13K, is shown in the charge cycle containing point (a) of Fig. 7A where the cyclical response of the charge curve is smooth. When the temperature maximum was about 17K a step appears in the downward side of the charge cycle at point (b) corresponding to the rising side of the temperature swing, but starting near the maximum value of the temperature in that particular cycle. The temperature maximum of the next temperature cycle was increased to about 20K and the step in the decreasing side of the charge cycle became a spike, point (c). The maximum temperature was raised for the next three thermal swings with the result that the spike on the decreasing side of the charge cycle became larger with each increase in the maximum temperature. The point (d) shows the largest spike in the charge corresponding to the largest swing in temperature, 4.2K to about 45K to 4.2K, in this experiment. Subsequent temperature swings through successively smaller ranges resulted in smaller spikes in the charge readings. Magnification of the three cycles indicated as points (a), (b) and (c) are shown in Fig. 7B. The charge cycle associated with the second thermal swing shows the step in the response very clearly. In the next thermal cycle, with a larger swing in temperature, the step becomes a well-defined spike. There is no obvious change in the charge response for the first thermal cycle. The appearance of the spikes suggests that at a certain temperature above 4.2K, the negative polarization charge at the –z base decreases numerically giving a positive response of the electrometer reading of the change in the polarization charge with respect to the polarization charge of the crystal at 4.2K[14]. The cyclical fluctuations, on which the spikes are riding, are attributed to pick up signals associated with the motion of the crystal assembly. It is noted that the amplitudes of the charge cycles show a small increase accompanying the increase in the temperature swings. Further discussion of the background effects will appear below.

It is seen that the determination of the upper limit to the temperature of the exit of the crystals from “saturation” of the polarization charge resulted in lowered values as the sensitivity of the measuring apparatus was increased. For example the data used to plot Fig. 3 indicated an exit temperature below about 36K, whereas the more sensitive scale used to accumulate the data of Figs. 4-7 indicated an upper limit of 15K for the crystals. A still more sensitive arrangement was employed to obtain a better determination of the exit temperature of the crystals. A series of measurements with the electrometer in the current mode, scale: $-0.50\text{x}10^{-12}$amp to $+0.50\text{x}10^{-12}$amp, were undertaken. As long as the polarization charge remained constant the current reading of the electrometer read zero. When the polarization charge changed the current became something other than zero, the larger the change in charge per unit time the larger the current reading. Starting at a crystal temperature of 4.2K the temperature of the crystal assembly was raised slowly by raising the crystal assembly out of the liquid helium. The results of the current measurements on the –z base of a crystal of $LiNbO_3$ are presented in Figs. 8A and 8B. The plot of current versus time shows a pervading cyclical fluctuation, which accompanies the motion of the crystal assembly. This fluctuation is attributed to pick-up from the motion of the magnets moving with the crystal assembly and the varying currents in the temperature sensor circuits. Further discussion of this attribute of the experiment will appear below. It is seen in Fig. 8A, that for temperature swings from 4.2K to about 17K, a slight reversal of the direction of the current pattern occurs near the valley of the pick-up cycle, or just before the maximum temperature is reached. The direction of the current is now positive and numerically increasing. At the summit of the temperature change, about 17K, the current momentarily stops. As the temperature decreases the current swings negative to combine with the positive current of the pick-up cycle. Point (a) of Fig. 8A illustrates this behavior for the temperature swing of 4.2K to 17K. As illustrated by point (b), the change in direction of the current almost disappears for the temperature swing of 4.2K to about 15K. The results of temperature swings of 4.2K to about 21K are indicated at point (c) and the cycles that follow it. It is seen that the positive current changes direction by increasing numerically on the down side of the positive pick-up current. As the temperature increased the current increased because the rate of change of the polarization charge relative to that at 4.2K increased as is evident in Fig. 3. When the temperature stopped changing the current stopped and the electrometer went to zero. As the temperature started to decrease from 21K, the current was large but negative thereby decreasing the positive current of the background. The large excursion of the polarization current, as seen at point (c), during the large temperature swing, was recognized as the signature response of the electrometer in the current mode to the exit of the crystal from "saturation". Therefore, the change in direction of the current at point (a) was recognized as the sign of the exit from "saturation". It is observed that the ripples at the top of every current cycle show no change in amplitude as the temperature swings increase in amplitude. It is concluded, therefore, that the ripples at the top of the current cycles are not manifestations of the changes in the polarization charge. A magnification of the three cycles about point (b) are presented in Fig. 8B. Based on the deductions made from Fig. 8A it is seen that in Fig.8B the temperature swing of the crystal just barely exceeds the exit temperature from "saturation". A similar set of data was obtained for crystals of $LiTaO_3$. The measurement on crystals of $CsNO_3$ appeared to have some minor differences from crystals of $LiTaO_3$ and $LiNbO_3$. These differences are attributed to the mosaic structure of the home-grown crystals of $CsNO_3$. The polarization current and corresponding temperature swings of a $CsNO_3$ crystal are presented in Fig. 9 as functions of time. The response, indicating the emergence from "saturation", is unambiguously identified at point (a). The large peak, point (a), corresponds to the comparatively large temperature swing of 4.2K to about 11K. Guided by this assignment of the part of the current response that comes from the emergence from "saturation", point (b) is recognized as an attribute of the exit from "saturation" for smaller temperature swings of 4.2K to about 7K. Points such as point (c) appear in every peak without much change in size as the temperature swings change amplitudes, and are considered to be background caused by the irregular cyclical motion of the crystal assembly.

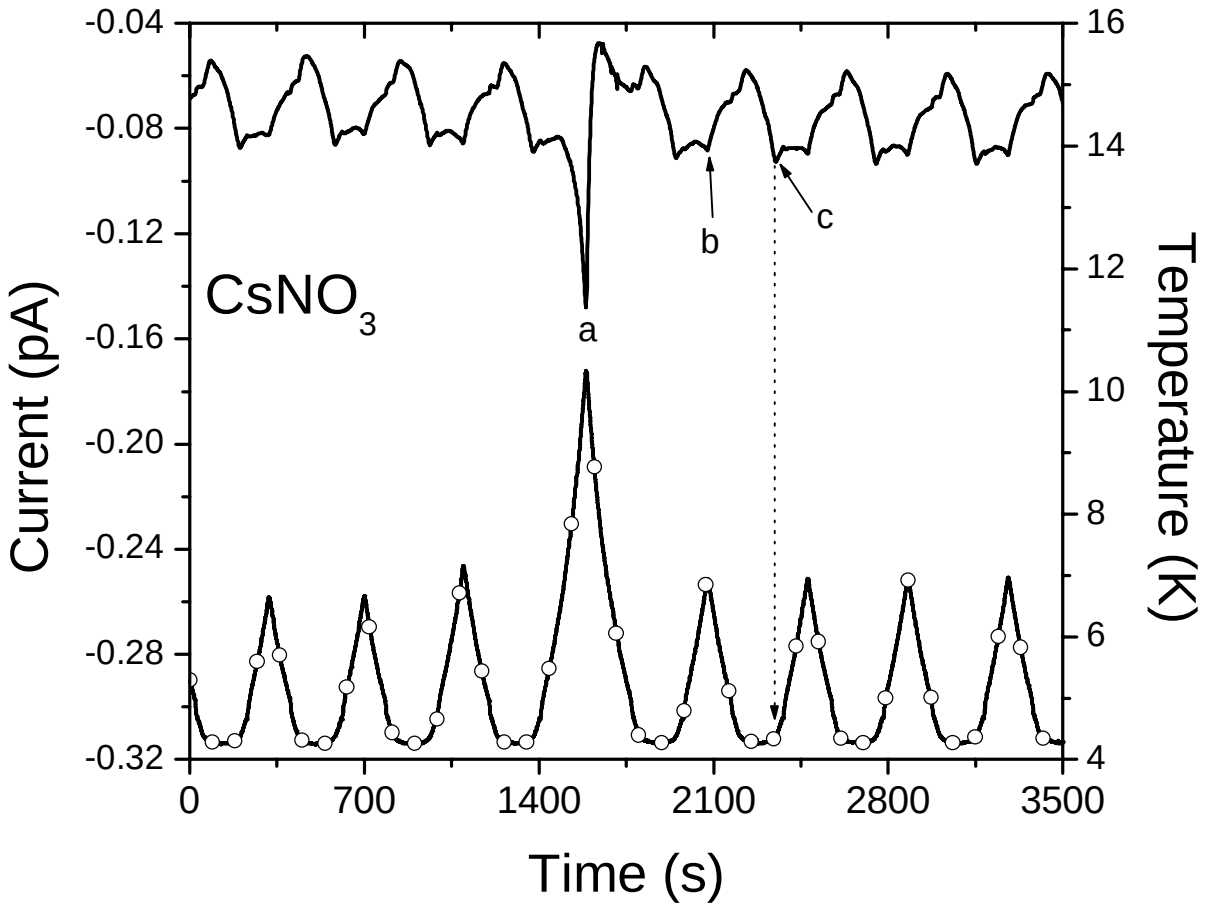

Figure 9: Temperature and polarization current at the +z base of a crystal of $CsNO_3$ as functions of time. The curve with the open circles represents the temperature as a function of time. The solid curve represents the current, consisting of the polarization current and the background current, as a function of time. Point (a) is the telltale peak, which identifies the polarization current, as the temperature of the crystal exceeds the exit temperature, i.e., the rate of change of the decreasing amount of positive polarization charge increases as the temperature increases above the exit temperature. Guided by the identification of the peak at point (a), point (b) can be attributed to the appearance of the polarization current at a temperature just above the exit temperature. Point (c) is a peak, which appears in the plots of the current for all temperature cycles and is identified as part of the background current.

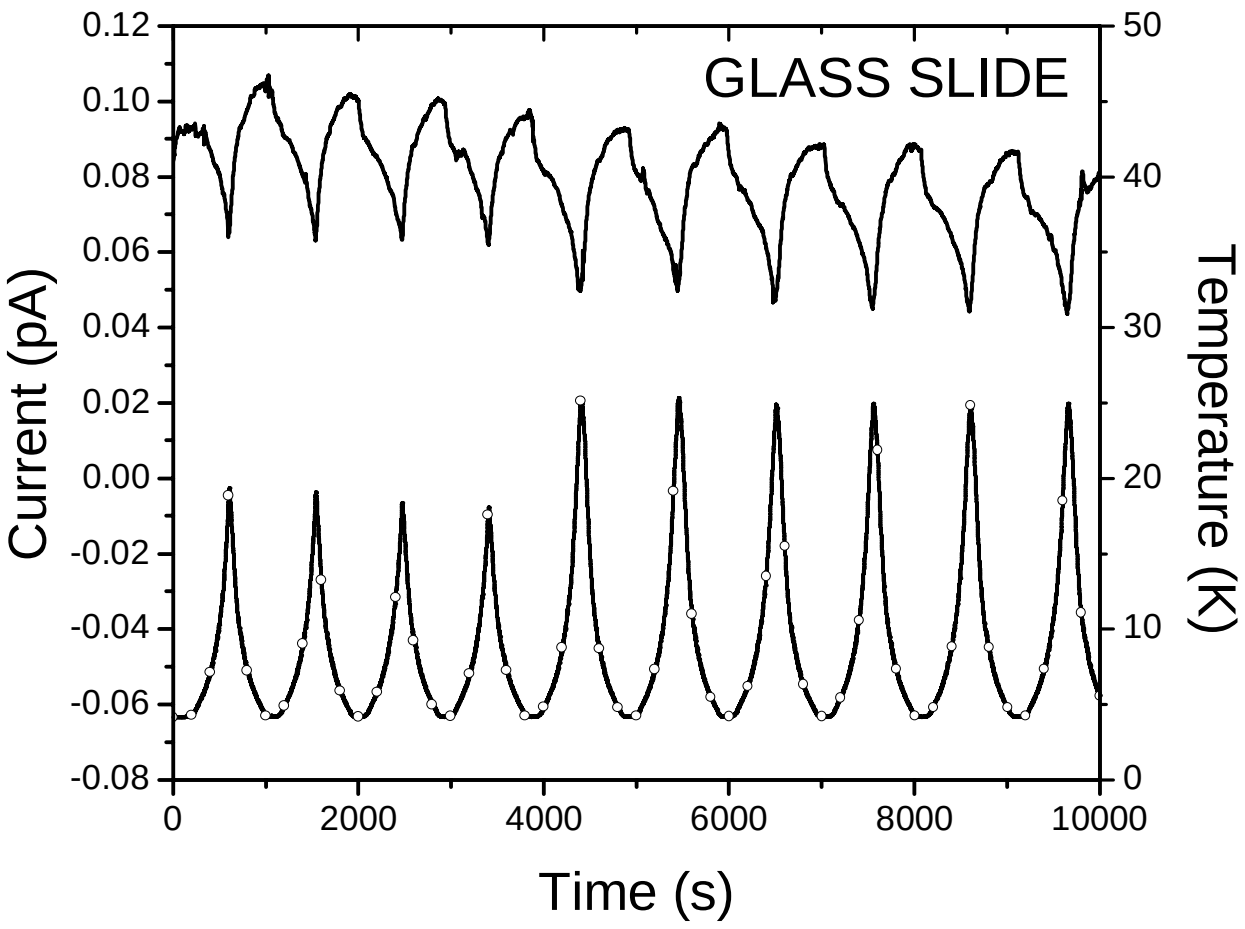

Figure 10A: Response of glass slide to cycles of temperature from 4.2K up to 25K. The curve with the open circles represents the temperature of the glass slide as a function of time. The solid curve is the current as a function of time as the temperature changed.

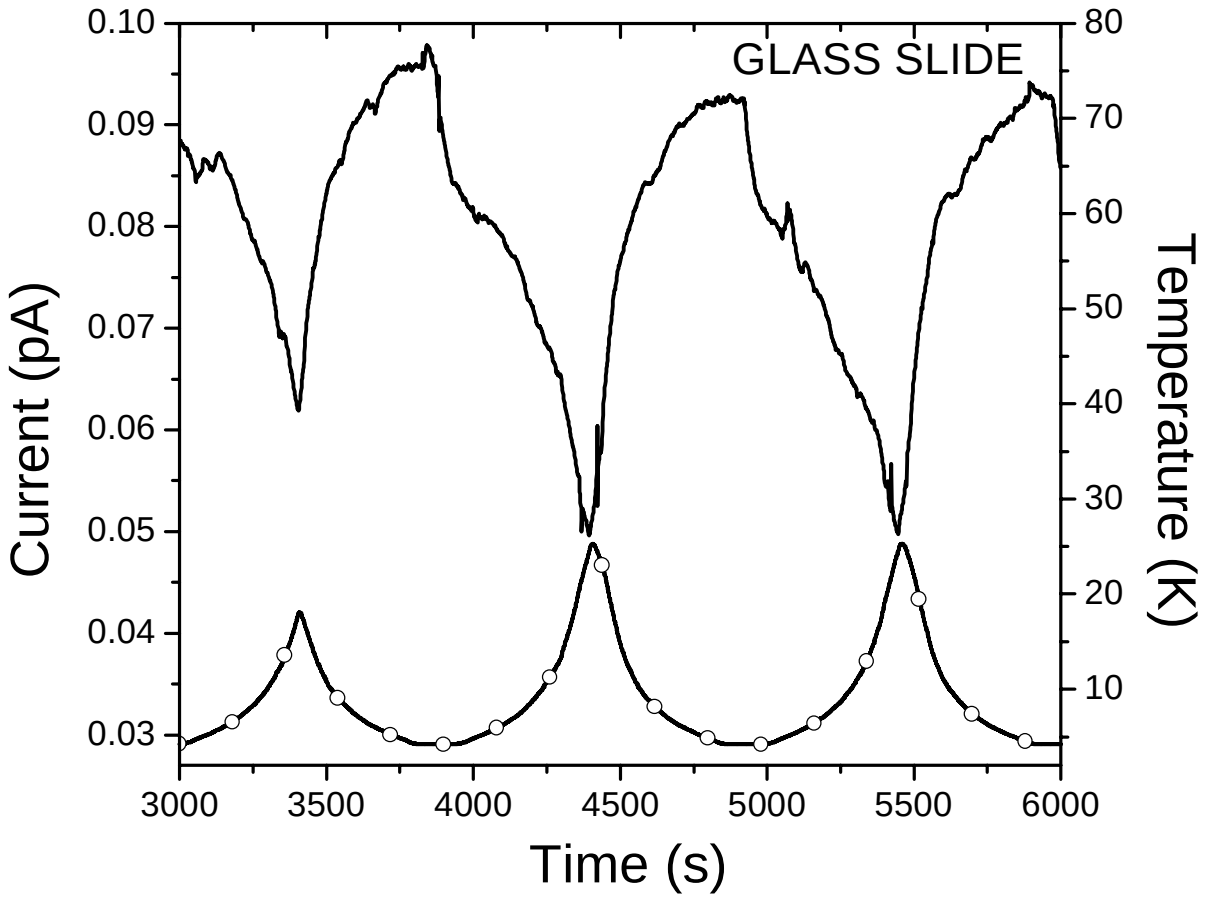

Figure 10B: Amplification of three temperature swings in Fig.11A.

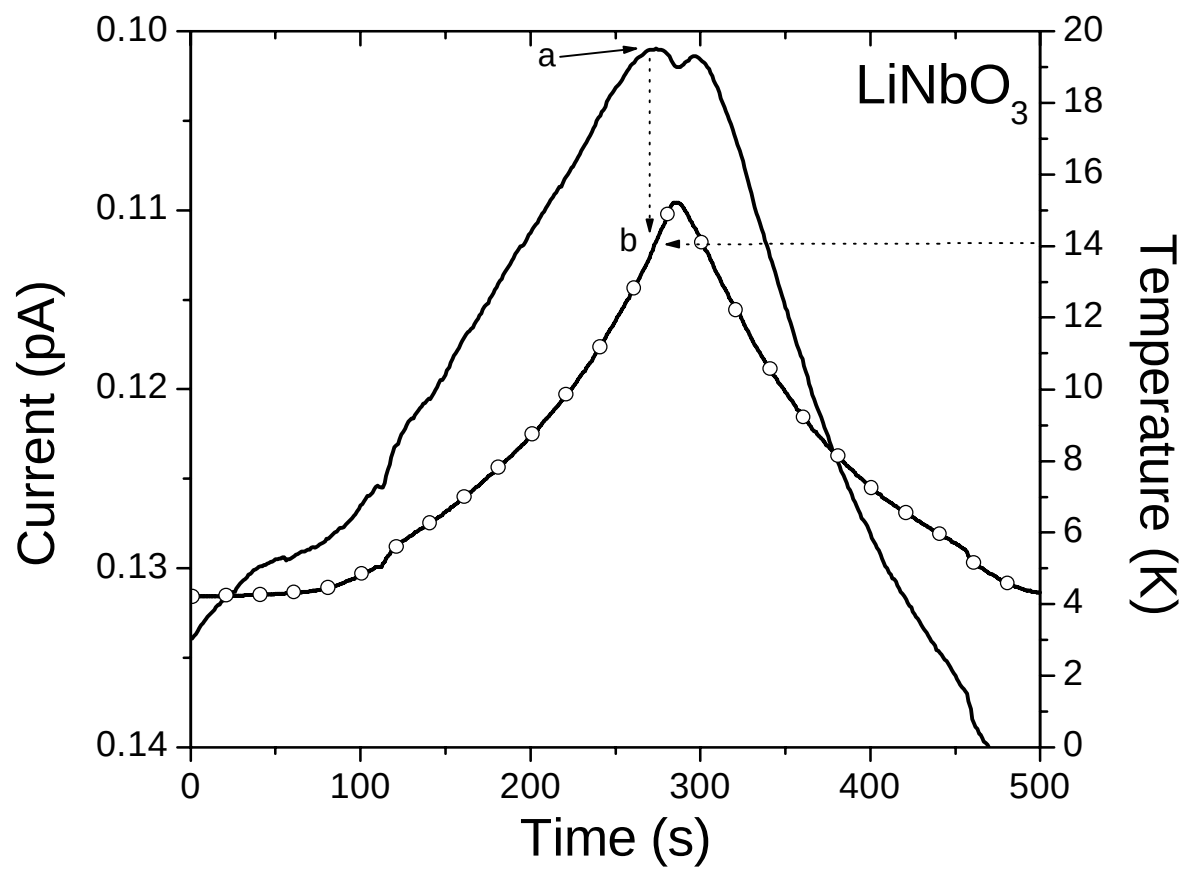

Figure 11: Illustration of the method used to determine the temperature of the exit from "saturation" of the crystal of LiNbO3 as applied to one of the temperature excursions in Fig. 8A. The curve with the open circles is the plot of the temperature as a function of time. The solid curve unmarked is the plot of the total current as a function of time. The ordinate for the plot of the polarization current is increasing downwards for presentation purposes. At point (a) on the current curve the current appears to be changing curvature as the polarization current starts to build up in the direction opposite to the background current. This happens at point (b) on the temperature curve. The data is then read for the value of the temperature corresponding to point (b).

It is seen from Figs. 7-9 and from other data that there is a cyclical component to the charge measurements and the current measurements, which occurs for small temperature swings from 4.2K to temperatures below the exit points as well as for larger temperature swings that exceed the exit temperature for each of the types of crystals investigated. In order to understand the origin of this cyclical fluctuation, a glass slide was substituted for the crystal in the crystal assembly and the apparatus was operated to reproduce the temperature cycles of the actual measurements. This was done to parallel the measurements of the charge and the measurements of the current. The results of the current measurements are presented in Figs.10A and 10B. In Fig.10A cyclical current fluctuations appear which accompany the motion of the slide assembly in and out of the liquid helium. As the amplitude of the temperature swing increases the amplitude of the current fluctuation increases. The current fluctuations are attributed to "pick-up" by the electrometer circuit caused in part by the motion of the magnets used to trigger the limiting mechanisms of the assembly. The cyclical fluctuations of the "pick-up" current are seen to be riding on a slowly varying background, which is attributed to the environment. Hereafter, the sum of the two effects described here will be called the background current. Three successive cycles of the fluctuations of the background current are shown in a magnified picture, Fig 10B. The first cycle results from a temperature swing with amplitude smaller than that of the following two swings. The amplitudes of the resulting current fluctuations follow the behavior of the temperature swings. In this magnified presentation, the current response shows some jagged parts, which reflect the irregular motion of the assembly.

The scheme, by which the temperature of the exit from "saturation" was determined, is illustrated in Fig.11. The temperature of the crystal of $LiNbO_3$ and the current are plotted as functions of time. For illustrative purposes, the curve representing the current is reversed, i.e., the scale of the current is decreasing upwards. This particular temperature swing is barely large enough to allow the crystal to emerge from "saturation". At first, at liquid helium temperature and slightly above 4.2K, the current consists only of the background current. The polarization charge is constant or almost constant and does not contribute to the current. The positive background current is decreasing when the exit from "saturation" initiates the polarization current. The total current is then a combination of the background current and the positive polarization current. At first this emergence from "saturation" will show a slowing down of the rate of decrease of the total current. Ultimately the polarization current will be large enough to change the direction of the change in total current, i.e., the total current will appear to be increasing. The polarization current goes to zero when the temperature swing reaches its maximum. During the temporary halt at the temperature maximum the current, which is only the background current, resumes it's decreasing path. As the temperature decreases from its maximum the polarization current resumes, but in a negative direction. Meanwhile the background current, responding to the motion of the crystal assembly, starts to increase. The total current is the sum of the increasing positive background current and the numerically decreasing negative polarization current until the temperature falls to the point of the onset of "saturation". From this point on the total current is solely background current. The determination of the point where the rate of decrease of the positive current slows down noticeably, point (a), and is followed by an increase in the positive current is noted in Fig. 11. This point is recognized as the exit point from saturation. The corresponding temperature, point (b), is also indicated. A large number of temperature cycles, each one just barely exceeding the exit point from "saturation" for the different types of crystal, were studied to determine the exit points The results obtained for a large number of thermal cycles for each crystal type are presented in Tab. I.

Table I: Temperatures at which the polarization charge of the crystals departs from constant values, i.e., from the "saturation" of the polarization charge of the crystals. This temperature is denoted as the "exit temperature".

| | Exit Temperature (K) | Number of Thermal Cycles | Area of z base ($mm^2$) | Thickness (mm) |
|---|---|---|---|---|
| $LiNbO_3$ | 14.09± 0.05 | 55 | 10.19± 0.04 | 1.001± 0.001 |
| $LiTaO_3$ | 11.384± 0.011 | 160 | 10.54± 0.07 | 1.00±0.00 |
| $CsNO_3$ | 6.461± 0.011 | 125 | 9.312± 0.021 | 3.012± 0.001 |

## CONCLUSION

The measurements of the polarization current, which result from changes in the polarization charge caused by temperature changes of the pyroelectric crystals, proved to be more sensitive than measurements of the change in the polarization charge for the determination of the temperature of the exit point from "saturation" of the polarization charge of the crystals. The sensitivity of the charge measurement was of the order of hundredths of nanocoulombs whereas the sensitivity of the current measurements was of the order of hundredths of picoamperes, i.e., hundredths of picocoulombs per second.

The peak in the current plot, superimposed on the background current, was identified as the polarization current by reference to the large peaks in Fig. 8A (point (c)) and in Fig. 9 (point (a)). The actual determination of the exit points was made for a large number of temperature swings from 4.2K to temperatures just above the point of appearance of the polarization current on top of the background currents. It was found that the exit temperatures from "saturation" were (14.09±0.05)K, (11.384±0.011)K and (6.461±0.011)K for $LiNbO_3$, $LiTaO_3$, and $CsNO_3$ respectively. The results are given in Table I.

*jdbjdb@binghamton.edu